\documentclass[
  aps,
  pre,
  reprint,
  superscriptaddress,
  nofootinbib,
  longbibliography,
  numerical,
  floatfix
]{revtex4-2}

\usepackage{amsmath}
\usepackage{amssymb}
\usepackage{graphicx}
\usepackage{xcolor}
\usepackage{derivative}
\usepackage{booktabs}
\usepackage{siunitx}
\usepackage{microtype}
\usepackage{tikz} 
\usepackage{pgfplots}
\usepgfplotslibrary{groupplots}
\pgfplotsset{compat=1.18}
\usepackage[hidelinks]{hyperref}
\hypersetup{
  colorlinks=true,
  linkcolor=blue,
  urlcolor=blue,
  citecolor=blue
}
\usepackage{cleveref}

\graphicspath{{./figs/}{./figs_ab/}}

\newcommand{\rv}{\mathbf{r}}
\newcommand{\nvac}{n_{\mathrm{vac}}}

\newcommand{\ctwo}{c^{(2)}}
\newcommand{\cone}{c^{(1)}}
\newcommand{\Fex}{F_{\mathrm{ex}}}

\newcommand{\Amin}{A_{\min}}
\newcommand{\Deltafree}{\Delta_{\alpha a}}
\newcommand{\Pins}{P_\mathrm{ins}}

\DeclareMathOperator\erf{erf}

\begin{document}

\title{
Free-volume origin of diverging direct correlations in hard crystals: insights from an exact one-dimensional model}

\author{Alessandro Simon}
\email{alessandro-rodolfo.simon@uni-tuebingen.de}
\affiliation{Institute of Applied Physics, University of T\"ubingen, Germany}

\author{Martin Oettel}
\affiliation{Institute of Applied Physics, University of T\"ubingen, Germany}
\email{martin.oettel@uni-tuebingen.de}

\date{\today}

\begin{abstract}
Second order direct correlation functions in three-dimensional hard-sphere crystals have much larger amplitudes (of the order of $1/n_{\mathrm{vac}}$ where $n_{\mathrm{vac}}$ is the vacancy concentration) and a more strongly structured spatial form of apparent shorter range than their liquid-state counterparts. We separate the two underlying questions---why the correlations are large and why they have the form they do---using the exact one-dimensional Percus functional for hard rods. A periodic crystal-like density is represented by Gaussian peaks with occupation probability $q=1-n_{\mathrm{vac}}$. The relevant thermodynamic quantity is the local insertion free volume $A(x)$.
Its minimum can be written as
$A_{\mathrm{min}} =\Delta_{\alpha a}+(1-\Delta_{\alpha a})n_{\mathrm{vac}}$, where $\Delta_{\alpha a}$ is the residual free volume at full occupation caused by finite localization (characterized by a width parameter $\alpha$) and lattice spacing $a$. The first and second direct correlations therefore contain the singular dependences $c^{(1)}\sim\ln A_{\mathrm{min}} $ and $c^{(2)} \sim-1/A_{\mathrm{min}} $. The observed $1/n_{\mathrm{vac}}$ dependence is the vacancy-dominated limit $n_{\mathrm{vac}}\gg\Delta_{\alpha a}$, rather than the most general result. The spatial form of $c^{(2)}$ has a separate geometrical origin: the hard-rod weight functions select configurations in which exclusion intervals and their boundaries intersect regions of small free volume. This produces plateaus, edges, and localized ridges tied to the underlying periodic density. A numerical solution of the inhomogeneous Ornstein--Zernike equation shows how these singular direct correlations are redistributed in the total and pair correlations. The model provides a minimal free-volume explanation for both the magnitude and the lattice-specific form of crystalline direct correlation functions.
\end{abstract}

\maketitle

\section{Introduction}

Hard spheres crystallize at sufficiently high pressure/packing fractions despite having a purely repulsive interaction \cite{royall2024colloidal}. In classical density functional theory (cDFT), crystals appear as free energy minima belonging to periodic one-body density distributions. 
In combination with fundamental-measure-theory (FMT) functionals, cDFT gives an accurate description of hard-sphere fluids and crystals \cite{evans1979nature,roth2010fundamental,hansen2006density,oettel2010free,PhysRevLett.84.694,PhysRevE.102.062137}.

The excess free-energy functional also generates the hierarchy of direct correlation functions,
\begin{equation}
  c^{(n)}(\rv_1,\ldots,\rv_n)
  =
  -\beta
  \frac{\delta^n \Fex[\rho]}
       {\delta\rho(\rv_1)\cdots\delta\rho(\rv_n)}.
\end{equation}
of which the second order function is of particular interest as it describes the (direct) correlation between two particles (it will be called DCF in short).
In a homogeneous liquid, translational invariance reduces the DCF to a function of one separation. In a crystal, it depends on two positions separately and inherits the periodic inhomogeneity of the density.

Recent FMT calculations for hard-sphere crystals show that the crystalline DCF differs qualitatively from its liquid-state counterpart \cite{lin2021direct,simon2026density}. Its amplitude is much larger, its dominant features are spatially concentrated to separations of the two positions smaller than the hard sphere diameter, and its shape contains plateaus, edges, and dips tied to the crystal geometry. These observations raise two distinct questions:
\begin{enumerate}
  \item Why does $\ctwo$ become so large in a crystal, in particular why is its magnitude $\propto 1/\nvac$ where $\nvac$ is the vacancy concentration in the solid?
  \item Why does the large contribution have a shorter range compared to a liquid and why does it have its particular lattice-related form?
\end{enumerate}
A three-dimensional crystal combines strong localization, small vacancy concentrations, and nontrivial geometry, which makes these mechanisms difficult to separate directly.

Here we isolate them in the exactly solvable one-dimensional hard-rod system. There is no thermodynamic crystallization transition in one dimension, so we impose a periodic crystal-like density profile. This is deliberate: the aim is not to model the transition itself, but to expose the mechanism by which an already localized density produces large and structured functional derivatives.
This work is one of several recent studies that use exactly solvable one-dimensional models to investigate broader theoretical problems~\cite{bjbt-61d9,gcyg-yw98,PhysRevE.110.014702,Mizani16072026}.

The main result is that the two questions have complementary answers. The amplitude is controlled by the minimum local insertion free volume,
\begin{equation}
  A(x)=1-n_1(x),
\end{equation}
where $n_1(x)$ is one of the FMT weighted densities describing a local packing fraction.
The shape is controlled by the geometry of the FMT weight functions. When $A(x)$ becomes small, the free-energy curvature contains inverse powers of $A$. The large terms occur only where the exclusion intervals associated with two positions intersect the narrow boundary layers of the low-free-volume cavities. In the sharp, vacancy-dominated limit, $A_{\min}\sim\nvac$ and therefore $\ctwo\sim-1/\nvac$.

\section{Hard-rod model and local free volume}
\label{sec:model}

\subsection{Exact functional}

We consider hard rods of length $\sigma=1$ and radius $R=\sigma/2=1/2$
and analyze them using classical DFT. In classical DFT, equilibrium properties are derived from the grand canonical functional which for 1D systems has the general form
\begin{equation}
    \Omega[\rho] = F_\text{id}[\rho] + F_\text{ex}[\rho] - \int \odif{x} [\mu - V^\text{ext}(x) ]\rho(x)  \;. 
\end{equation}
Here, $\mu$ is the chemical potential and $V^\text{ext}(x)$ an arbitrary one-body external potential, acting on the centers of the rods. Furthermore, $ \beta F_\text{id}[\rho]= \int dx\, \rho(x)[ \ln (\rho \lambda) -1]$ is the ideal gas functional, with $\beta=1/(k_\text{B}T)$ the inverse temperature and $\lambda$ the thermal de Broglie length. For hard rods with diameter $\sigma=2R$, the exact excess functional $F_\text{ex}[\rho]$ due to Percus is given by~\cite{Percus1976HardRods}
\begin{align}
  \label{eq:percus-functional}
  \beta \Fex[\rho]
  &=
  \int \odif{x}\,\Phi\bigl(n_0(x),n_1(x)\bigr),\\
  \Phi(n_0,n_1)
  &=-n_0\ln(1-n_1).
\end{align}
The weighted densities are
\begin{equation}
  n_i(x)=\int \odif{x'}\,\rho(x')w_i(x-x'),
\end{equation}
with weight functions
\begin{align}
  w_0(x)&=\frac{1}{2}\delta(R-|x|),\\
  w_1(x)&=\Theta(R-|x|).
\end{align}
The weighted density $n_1(x)$ has the meaning of a local packing fraction and therefore the quantity
\begin{equation}
  \label{eq:free-volume-definition}
  A(x)=1-n_1(x)
\end{equation}
will be called the local insertion free volume. It is the central quantity in the considerations below.

The density profile in equilibrium ($\rho_\text{eq}$) minimizes the grand potential and thus fulfills
\begin{equation}
    \ln (\rho_\text{eq}(x) \lambda) - c^{(1)}(x) - \beta \mu + \beta V^\text{ext}(x) =0
    \label{eq:ele}
\end{equation}
where $c^{(1)}(x)= -\beta \delta F_\text{ex}/\delta \rho(x)|_{\rho=\rho_\text{eq}}$.

\begin{figure}[t]
  \centering
  \includegraphics[width=\linewidth]{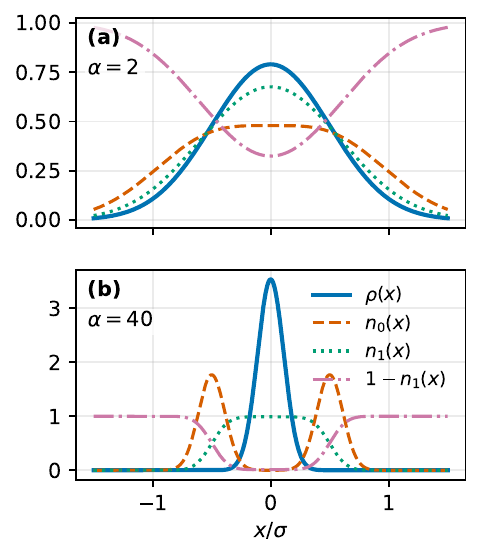}
  \caption{Density profiles and corresponding weighted densities for a broad peak (a) and a sharply localized peak (b) for $\nvac=\num{1e-2}$. The weighted packing fraction $n_1$ determines the local insertion free volume $A=1-n_1$. }
  \label{fig:density-weighted-densities}
\end{figure}

\subsection{Periodic Gaussian density}

Hard rods do not crystallize in 1D but a periodic, crystal-like density profile can be imposed by a suitable periodic external potential which is obtained by solving \Cref{eq:ele} for $V^\text{ext}(x)$, assuming a specific form for $\rho_\text{eq}$.  
Here we represent a crystal-like density by a periodic array of normalized Gaussian peaks,
motivated by the approximate Gaussian shape of the density peaks in the 3D hard sphere crystal \cite{oettel2010free},
\begin{equation}
  \label{eq:periodic-density}
  \rho_\text{eq}(x) = \rho_{q\alpha a}(x)
  =
  q\sum_{j\in\mathbb{Z}}\psi_\alpha(x-ja),
  \qquad
  q=1-\nvac,
\end{equation}
where 
\begin{equation}
  \label{eq:gaussian}
  \psi_\alpha(x)
  =
  \sqrt{\frac{\alpha}{\pi}}\exp(-\alpha x^2).
\end{equation}
We have labeled the crystal-like profile $\rho_{q\alpha a}(x)$ by $q$, the mean occupancy of a unit cell; $\alpha$, the width parameter of the Gaussian peak and $a$, the lattice spacing, to indicate its parametric dependence on these variables.
The mean occupancy and the lattice spacing are distinct sources of the effective free volume at a lattice site whose effect will be disentangled below.

When the influence of neighboring peaks on the density distribution inside a unit cell are negligible $(a>\sigma \gg 1/\sqrt{\alpha})$, the periodic problem reduces locally to a single Gaussian peak, 
\begin{equation}
  \rho_{q\alpha}(x)=q\psi_\alpha(x).
\end{equation}
For this profile the weighted densities are
\begin{align}
  \label{eq:n0-single}
  n_{0,q\alpha}(x)
  &=
  \frac{q}{2}\sqrt{\frac{\alpha}{\pi}}
  \left[
    e^{-\alpha(x+R)^2}+e^{-\alpha(x-R)^2}
  \right],\\
  \label{eq:n1-single}
  n_{1,q\alpha}(x)
  &=
  \frac{q}{2}
  \left[
    \erf\!\bigl(\sqrt{\alpha}(x+R)\bigr)
    -
    \erf\!\bigl(\sqrt{\alpha}(x-R)\bigr)
  \right].
\end{align}
Broad and sharply localized examples of $\rho_{q\alpha}$ and the corresponding weighted densities, as well as the local free volume are shown in \Cref{fig:density-weighted-densities} (for $q=0.99$). It can be seen that for the sharply localized peak there is a strip of finite width where the local free volume is $1-q\approx 0$, in contrast to 
 the local free volume of the broad peak which shows a broad dip with a minimum well above zero.

\begin{figure}[t]
  \centering

  \includegraphics[width=1\linewidth]{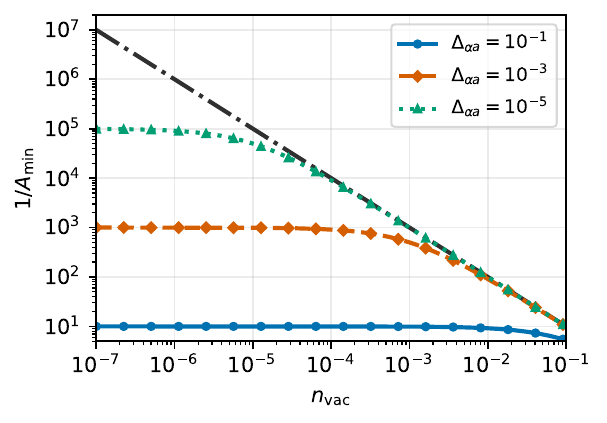}
  \caption{Crossover for the inverse minimal free volume  implied by \Cref{eq:Amin-crossover}. The minimum insertion free volume follows $\Amin\simeq\nvac$ only while the vacancy contribution exceeds the residual free volume $\Deltafree$. The corresponding inverse-free-volume amplification saturates at $1/\Deltafree$ when $\nvac\ll\Deltafree$. }
  \label{fig:free-volume-crossover}
\end{figure}

\subsection{Localization--vacancy crossover}
\label{sec:crossover}

Because the density profile in \Cref{eq:periodic-density} is proportional to the site
occupation $q$, the weighted densities inherit the same linear dependence.
We therefore write
\begin{equation}
    n_{1,q\alpha a}(x)=q\,\bar n_{1,\alpha a}(x),
\end{equation}
where $\bar n_{1,\alpha a}(x)$ is the weighted packing fraction for a fully occupied
lattice. Its maximum,
\begin{equation}
    m_{\alpha a}=\max_x \bar n_{1,\alpha a}(x),
\end{equation}
determines the smallest local free volume in the fully occupied lattice which we denote by
\begin{equation}
    \Delta_{\alpha a}=1-m_{\alpha a},
\end{equation}
It can be zero if $\alpha\to \infty$ ($\delta$-peak) or $a=\sigma$.
The minimum
of the local free volume $A_{q\alpha a}(x)=1-n_{1,q\alpha a}(x)$ in the lattice with vacancies is then given by
\begin{align}
    A_{\min}
    &=1-qm_{\alpha a} \\
    &=\Delta_{\alpha a}
      +\left(1-\Delta_{\alpha a}\right)\nvac,
      \label{eq:Amin-crossover}
\end{align}
where $q=1-\nvac$. %

The two terms in this expression have different origins. The quantity
$\Delta_{\alpha a}$ is the residual free volume caused by the finite width
of the density peaks and the lattice spacing which determines the overlap between
neighboring peaks. The second term is the additional free volume produced
by vacancies. Their relative magnitude determines which mechanism controls
the minimum free volume:
\begin{align}
    \nvac \gg \Delta_{\alpha a}
    &\quad\Longrightarrow\quad
    A_{\min}\simeq \nvac, \\
    \nvac \ll \Delta_{\alpha a}
    &\quad\Longrightarrow\quad
    A_{\min}\simeq \Delta_{\alpha a}.
\end{align}
Thus, decreasing the vacancy concentration does not necessarily make the
free volume vanish. For a density profile of fixed finite width,
$A_{\min}$ eventually saturates at $\Delta_{\alpha a}$. The
vacancy-controlled singularity appears only when the residual free volume
is smaller than the vacancy contribution.
A schematic plot of the crossover behavior in the inverse minimal free volume is shown in \Cref{fig:free-volume-crossover}.

For an isolated Gaussian peak, the maximum weighted packing fraction is
\begin{equation}
    m_{\alpha}=\erf\!\left(\sqrt{\alpha}\,R\right),
\end{equation}
and hence
\begin{equation}
    \Delta_{\alpha}
    =1-\erf\!\left(\sqrt{\alpha}\,R\right).
\end{equation}
The order of limits therefore matters. Taking $\alpha\to\infty$ first gives
$\Delta_{\alpha}\to0$, so that $A_{\min}\sim\nvac$. By contrast, taking
$\nvac\to0$ at fixed $\alpha$ gives
$A_{\min}\to\Delta_{\alpha}$, and %
a possible divergence upon $\nvac \to 0$
is cut off.

\section{The large magnitude of the crystal direct correlation function }
\label{sec:amplitude}

\subsection{First order direct correlation function}

The first order direct correlation function for a density profile with local free volume $A(x)$ and
weighted density $n_0(x)$ is given by
\begin{multline}
  \label{eq:c1-exact}
  \cone(x)
  =
  \frac{1}{2}
  \left[
    \ln A(x+R)+\ln A(x-R)
  \right]
  \\
  -
  \int_{x-R}^{x+R}\odif{s}\,
  \frac{n_0(s)}{A(s)}.
\end{multline}
The Widom insertion relation~\cite{widom1963topics} %
\begin{equation}
  \cone(x)=\ln P_{\mathrm{ins}}(x),
\end{equation}
relates $\cone$ to $P_{\mathrm{ins}}(x)$ which is the inhomogeneous (local) insertion probability for a hard rod with center at $x$. Hence a small local insertion probability produces a large negative $\cone$.

\begin{figure}[t]
  \centering
  \includegraphics[width=\linewidth]{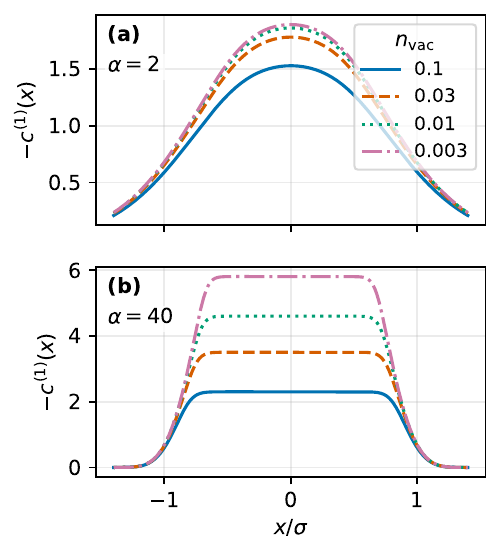}
  \caption{First direct correlation function for a single broad density peak (a) and a sharply localized peak (b). The strong vacancy dependence in panel (b) reflects the regime $\Amin\simeq\nvac$; the weak dependence in panel (a) reflects saturation at a finite residual free volume.}
  \label{fig:c1-vacancy-scan}
\end{figure}

The singular dependence due to small insertion probability follows directly from \Cref{eq:c1-exact}. 
The first two terms contain a dependence $\propto \ln \Amin$ whereas the integral term contains a $1/A$ term.
However, for a sharply localized peak $\rho(x)$ the weighted density $n_0$ is peaked at $x \pm R$, and near those points  
\begin{equation}
  n_0(s)\simeq\frac{1}{2}|A'(s)|.
\end{equation}
Consequently, the integral term in \Cref{eq:c1-exact} becomes
\begin{equation}
  \int\odif{s}\,\frac{n_0(s)}{A(s)}
  \sim
  \frac{1}{2}\int\frac{\odif{A}}{A}
  \sim -
  \frac{1}{2} \ln{\Amin}.
\end{equation}
Therefore,
\begin{equation}
  \cone_{\mathrm{sing}}\sim\ln\Amin.
  \label{eq:c1-scaling}
\end{equation}
This links the minimum insertion probability to the minimum local free volume, $\text{min} (P_\text{ins}) = \Amin$. \Cref{fig:c1-vacancy-scan}(b) shows $-\cone$ for an isolated sharp density peak and for various $\nvac$. It is plateau-like and the plateau height is $-\ln\nvac$. Thus the insertion probability $P_\text{ins}(x) \approx \nvac $ is also roughly constant on the plateau.
For broad peaks, $\Deltafree$ remains appreciable and the vacancy dependence is weak, see \Cref{fig:c1-vacancy-scan}(a). 

\subsection{Second order direct correlation function}

The second order direct correlation function is
\begin{eqnarray}
  -\ctwo(x,x')
  &=&
  \beta
  \fdv[style-var=multiple]{\Fex[\rho]}{\rho(x),\rho(x')}. \label{eq:c21d} \\ 
  &=& K_{01}(x,x')+K_{10}(x,x')+K_{11}(x,x') \nonumber
\end{eqnarray}
where
\begin{equation}
  \label{eq:Kij}
  K_{ij}(x,x')
  =
  \int\odif{s}\,
  \Phi_{ij}(s)w_i(s-x)w_j(s-x').
\end{equation}
and 
$\Phi_{ij}(s)= \pdv{\Phi(s)}/{n_i(s),n_j(s)}$
are second derivatives of the local free-energy density.  
The nonzero ones are given by
\begin{align}
  \Phi_{01}(s)=\Phi_{10}(s)&=\frac{1}{A(s)},\\
  \Phi_{11}(s)&=\frac{n_0(s)}{A(s)^2}.
\end{align}

Evaluating the convolution-type integrals in \Cref{eq:Kij} results in 
\begin{multline}
  \label{eq:c2-explicit}
  -\ctwo(x,x')
  =
  \frac{\chi_{[-1,0]}(x-x')}{2A(x+R)}
  +
  \frac{\chi_{[0,1]}(x-x')}{2A(x-R)}
  \\
  +
  \frac{\chi_{[0,1]}(x-x')}{2A(x'+R)}
  +
  \frac{\chi_{[-1,0]}(x-x')}{2A(x'-R)}
  \\
  +
  \chi_{[-1,1]}(x-x')
  \int_{\max(x-R,x'-R)}^{\min(x+R,x'+R)}
  \odif{s}\,
  \frac{n_0(s)}{A(s)^2}.
\end{multline}
Here $\chi_I$ is the indicator function of the interval $I$.

The first four terms in \Cref{eq:c2-explicit} sample $1/A$ directly and therefore scale as $1/\Amin$ whenever their shifted arguments enter a low-free-volume region. The last term contains 
a possible stronger divergence $n_0/A^2$ for small local free volume.
Using a similar argument as in the derivation of \Cref{eq:c1-scaling} for sharp peaks
($n_0 \simeq A'/2$) shows that the integral of the last term is $\propto 1/A$ (a detailed treatment is given in App.~\ref{app:K11})
and therefore all singular contributions have the same leading algebraic order,
\begin{equation}
  \ctwo_{\mathrm{sing}}\sim-\frac{1}{\Amin}.
  \label{eq:c2-Amin-scaling}
\end{equation}
The large crystalline amplitude is thus a free-energy-curvature effect caused by very small insertion free volume.

In the vacancy-dominated regime, \Cref{eq:c2-Amin-scaling} reduces to
\begin{equation}
  \ctwo_{\mathrm{sing}}\sim-\frac{1}{\nvac}.
  \label{eq:c2-nvac-scaling}
\end{equation}
At fixed finite localization, the divergence eventually crosses over to saturation at $-1/\Deltafree$.

\begin{figure*}
  \centering
  \includegraphics[width=\linewidth]{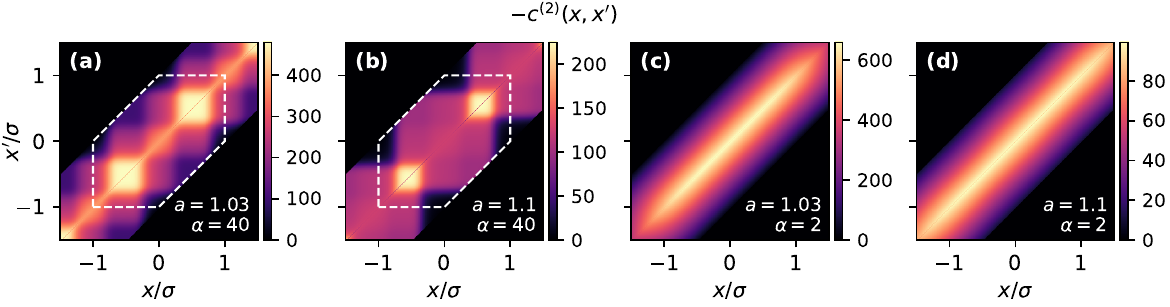}
  \caption{Full second direct correlation function for sharply localized peaks (a, b), and broad density peaks (c, d) at $\nvac = 1\times10^{-2}$.
  The lattice spacing was either $a/\sigma=\num{1.03}$ for (a, c) and $a/\sigma=\num{1.1}$ for (b, d). The strict support (marked by the dashed region) remains fixed by the rod length, but the dominant large-amplitude contribution becomes concentrated along geometrically selected regions in the sharp case. For the numerics, a lattice with five density peaks has been used, shown is only the region around the middle peak.}
  \label{fig:c2-heatmap}
\end{figure*}

\section{The extent of the crystal direct correlation function}
\label{sec:shape}

Having established the large magnitude of the direct correlation function, it is instructive to discuss the full function in the $x-x'$ plane, using the explicit form in \Cref{eq:c2-explicit}. The indicator functions there define the support of the function. For a general profile, this support is given by a diagonal strip in the $x-x'$ plane with horizontal width 2$\sigma$. For an isolated delta peak at the origin the support has a hexagonal shape with the origin in its center, see also \Cref{fig:sharp-domain} in App.~\ref{app:sharp-oz}. Thus a periodic assembly of delta peaks leads to a support which is a periodic arrangement of these hexagons along the diagonal strip. Thus, for a general, crystalline profile we expect an overall shape which shows hexagon-type modulations in the diagonal strip.

In \Cref{fig:c2-heatmap} the full DCF is shown for sharp peaks [$\alpha\sigma^2=40$, panels (a) and (b)] and for broad peaks [$\alpha \sigma^2=2$, panels (c) and (d)]. Furthermore, we differentiate between a unit cell just larger than the rod length [$a/\sigma=1.03$, panels (a) and (c)], and a somewhat larger one [$a/\sigma=1.1$, panels (b) and (d)]. For sharp peaks and the larger unit cell, the hexagonal support of the single delta peak is clearly visible [dashed lines in (b)], the DCF is nearly homogeneous there except for a strong enhancement at the diagonal edges (e.g. at $x=x'\approx 0.8 \sigma$). These can be viewed as ``interference effects'' with the next unit cell. It is also visible that for $x=0$, the range in $|x'|$ of the high-magnitude region is clearly less than $\sigma$, being approximately 0.8$\,\sigma$. This is a feature which has been observed in 3D \cite{lin2021direct} and is highly distinct from a liquid DCF. If for sharp peaks the unit cell length is reduced, the DCF in the hexagonal support [dashed lines in (a)] shows more interference features and larger inhomogeneities, but the finding of a high-magnitude region somewhat reduced in size remains. For broad peaks, the DCF is close to that of a liquid and shows little structure. This holds for both values of the unit cell length $a$ [panels (c) and (d)]. The largest magnitude here is determined by the term $K_{11}$ in \Cref{eq:c21d} whose magnitude for a liquid-like configuration is $ \propto 1/\bar{A}^2$ where $\bar{A}$ is the average free volume [which is 0.038 for panel (c) and 0.1 for panel (d)].

The reduced range of the DCF for highly localized peaks can be rationalized by analyzing the full expression in \Cref{eq:c2-explicit}. The individual terms amount to contributions $\propto 1/A(x)$ which are shifted by $\pm R$ and added up. (For an analysis of the individual contributions to the DCF for a single peak, see App.~\ref{app:dcf_peak}). The inverse free volume for a delta-like peak ($1/\alpha=0$) is $\propto \theta(R-|x|)$ and has a FWHM of $2R$. However, as $1/\alpha$ increases from zero, the FWHM is  substantially reduced (it even approaches zero for $\nvac \to 0$). This reduced width of  $1/A(x)$ is ``transported'' to the DCF.
We should however stress that the mathematical support of the hard-rod direct correlation function does not become shorter-ranged: it remains restricted by $|x-x'|\leq\sigma$.

\section{Consequences for total and pair correlations}
\label{sec:oz}

The inhomogeneous Ornstein--Zernike equation is
\begin{equation}
  \label{eq:oz}
  h(x,x')
  =
  \ctwo(x,x')
  +
  \int\odif{s}\,
  \ctwo(x,s)\rho(s)h(s,x').
\end{equation}
A large direct correlation does not imply that $h$ diverges uniformly. The integral equation redistributes the singular kernel and enforces the hard-core structure of the pair correlations.
For an isolated delta peak of the density, $h(x_1,x_2)=-1$ on the hexagonal support domain $\mathcal{D}$ of $\ctwo$ (see \Cref{fig:sharp-domain} in App.~\ref{app:sharp-oz}).  If $\mathcal{S}$ denotes the square of side length $2\sigma$ with the peak at its center, then the solution for $h$ on the domain $\mathcal{D}'= \mathcal{S} \backslash \mathcal{D}$ becomes $h(x_1,x_2)=(1-\nvac)/\nvac$ (see App.~\ref{app:sharp-oz}) and thus it exhibits a similar singularity as the DCF.

For a numerical solution for an array of density peaks with finite width, we consider a grid with spacing  $\Delta x$ where \Cref{eq:oz} becomes
\begin{equation}
  H=C+CDH,
\end{equation}
with
\begin{equation}
  D_{ij}=\Delta x\,\rho_i\delta_{ij}.
\end{equation}
The numerical solution is obtained from
\begin{equation}
  \label{eq:oz-matrix}
  (I-CD)H=C. 
\end{equation}

\begin{figure}[t]
  \centering
  \includegraphics[width=\linewidth]{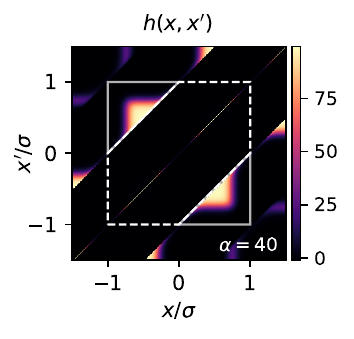}
  \caption{Numerical solution of the inhomogeneous Ornstein--Zernike equation for a sharply localized profile at $\nvac=\num{1e-2}$ and $a=\num{1.03}$. The dashed and full lines show the different support in the unit cell, cf.\ \Cref{fig:sharp-domain}.}
  \label{fig:h-heatmap}
\end{figure}

The comparison in \Cref{fig:h-heatmap} shows that the singular response is geometrically selective. At a density-peak center, exclusion largely fixes $h\simeq-1$ over the core region. Inside $\mathcal{D'}$, the OZ convolution can instead inherit the large inverse-free-volume contribution, however, similar to the reduced extent of $\ctwo$ (compared to $\mathcal{D}$)   the region of large $h$ is smaller than the support domain $\mathcal{D'}$ for the delta-like peak.

The two-body density is
\begin{equation}
  \label{eq:rho2}
  \rho^{(2)}(x,x')
  =
  \bigl[1+h(x,x')\bigr]\rho(x)\rho(x').
\end{equation}
It represents the physically weighted pair probability. Large values of $h$ near cavity boundaries are multiplied by small one-body densities whenever either coordinate lies between localized peaks. Consequently, the most singular normalized structures can have a much weaker manifestation in $\rho^{(2)}$.

\section{Discussion and conclusions}
\label{sec:discussion}

We used the exact one-dimensional hard-rod functional to answer two questions motivated by direct correlation functions in hard-sphere crystals.

First, $\ctwo$ is large because a crystal-like density creates regions of very small insertion free volume $A$. The exact Percus free-energy curvature contains $1/A$ and $n_0/A^2$, and the integrated singular kernel scales as
\begin{equation}
  \ctwo_{\mathrm{sing}}\sim-\frac{1}{\Amin}.
\end{equation}
For sharp density peaks around lattice sites, there is a vacancy-dominated regime where $\Amin \approx \nvac$ and thus the DCF shows an enhancement $\propto 1/\nvac$ which has been observed in 3D before. The physical explanation for this behavior is as follows: Consider first $\cone(x)= \ln P_\text{ins}(x)$ which is linked to the inhomogeneous insertion probability for an additional rod. Around the lattice site, $P_\text{ins}=\nvac$ since it reflects the probability to find an empty lattice site in the ensemble of crystal snapshots. Secondly, $\ctwo(x,x')=\delta\cone(x)/\delta\rho(x')$ describes the change of the logarithm of $P_\text{ins}(x)$ upon a density change at $x'$. If $x$ is a lattice point and $|x'-x|<\sigma$, then a positive density change $\delta\rho(x')$ is only possible if in some snapshots of the crystal ensemble a rod is inserted at $x'$, and this is only possible for snapshots where there is no rod at $x$ (i.e.\ in snapshots with vacancies). Therefore such an insertion lowers the insertion probability at lattice site $x$ by $\delta\Pins(x)= -\delta\rho(x')$ and $\ctwo(x,x') = - \partial \ln \Pins(x) / \partial \Pins(x) =- 1/\nvac$.

Second, the reduced range of high-magnitude regions of $\ctwo$ (compared to the liquid case) is a consequence of the geometry of the FMT weights. The interval weight determines whether two exclusion regions overlap, while the boundary weight determines whether the edge of one exclusion region enters a low-free-volume cavity. The large contribution is concentrated where these geometrical conditions coincide with the narrow boundary layers of $1/A(x)$. Periodicity then repeats the same structures according to the lattice geometry.

An analogous separation is built into higher-dimensional FMT: inverse local free-volume factors set the amplitude of the functional derivatives,
while convolutions of the scalar, vector, and tensor weights determine their spatial structure. The boundary points of the one-dimensional model
are replaced by particle surfaces and their overlap manifolds, producing geometry-dependent structures in crystalline DCFs.

The model has clear limitations. It does not describe a one-dimensional crystallization transition (of course), the density profile is imposed rather than found from unconstrained coexistence, and the Gaussian ansatz restricts the shape of the localized peaks. However, these limitations are also what make the mechanism transparent. The next test should be performed directly on three-dimensional hard-sphere crystal data by evaluating the minimum local free-volume measure and checking whether the dominant DCF amplitudes collapse against its inverse across vacancy concentration, localization strength, and lattice spacing.

\appendix

\section{Boundary-layer derivation of the $K_{11}$ scaling}
\label{app:K11}

The local curvature entering the final term of \Cref{eq:c2-explicit} is
\begin{equation}
  \Phi_{11}(s)=\frac{n_0(s)}{A(s)^2}.
\end{equation}
Pointwise, this can reach values of order $1/\Amin^2$. The convolution kernel, however, contains its integral through a boundary layer.

Near the right boundary of a sharply localized single Gaussian peak,
\begin{equation}
  n_1(s)
  \simeq
  q\int_{s-R}^{\infty}\odif{y}\,\psi_\alpha(y).
\end{equation}
Since $A=1-n_1$,
\begin{equation}
  A'(s)
  \simeq
  q\psi_\alpha(s-R),
\end{equation}
while
\begin{equation}
  n_0(s)
  \simeq
  \frac{q}{2}\psi_\alpha(s-R)
  \simeq
  \frac{1}{2}A'(s).
\end{equation}
Hence
\begin{align}
  \int\odif{s}\,\frac{n_0(s)}{A(s)^2}
  &\simeq
  \frac{1}{2}
  \int\odif{s}\,\frac{A'(s)}{A(s)^2}\\
  &=
  \frac{1}{2}
  \left(
    \frac{1}{A_{\mathrm{in}}}
    -
    \frac{1}{A_{\mathrm{out}}}
  \right).
\end{align}
With $A_{\mathrm{in}}\sim\Amin$ and $A_{\mathrm{out}}=O(1)$,
\begin{equation}
  K_{11}\sim\frac{1}{\Amin}.
\end{equation}
The same argument applies at the left boundary with the corresponding sign of $A'$. The absolute boundary contribution is positive in $K_{11}$, while $\ctwo$ carries the overall negative sign in \Cref{eq:c2-explicit}.

\section{Sharp-cavity Ornstein--Zernike limit}
\label{app:sharp-oz}

\begin{figure}[t]
  \centering
  \includegraphics[width=0.82\linewidth]{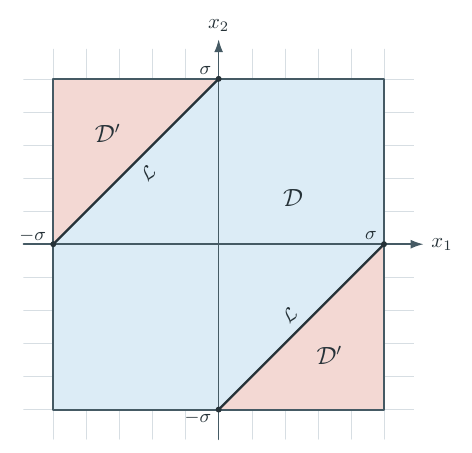}
  \caption{Schematic support region $\mathcal D$ of the limiting single-cavity direct correlation function and the boundary line $\mathcal L$ on which the singular total-correlation response appears.}
  \label{fig:sharp-domain}
\end{figure}

For a single infinitely sharp density peak,
\begin{equation}
  \rho(x)=q\delta(x),
  \qquad q=1-\nvac,
\end{equation}
and the direct correlation has a piecewise constant form, given by
\begin{equation}
  \ctwo(x_1,x_2)
  =
  \begin{cases}
    -1/\nvac, & (x_1,x_2)\in\mathcal D,\\
    0, & (x_1,x_2)\notin\mathcal D \;.
  \end{cases}
\end{equation}
The support region $\mathcal D$ and the boundary line $\mathcal L$ used below are shown schematically in \Cref{fig:sharp-domain}, as well as a region $\mathcal D'$ which completes $\mathcal D$ to a square.

The OZ equation reduces to
\begin{equation}
  h(x_1,x_2)
  =
  \ctwo(x_1,x_2)
  +q\ctwo(x_1,0)h(0,x_2).
\end{equation}
For $x_1,x_2$ inside region $\mathcal D$ where $\ctwo=-1/\nvac$ the equation becomes
\begin{equation}
    h(x_1,x_2) + \frac{1}{\nvac} = h(x_1,0) \frac{\nvac - 1}{\nvac} \;,
\end{equation}
and it is seen that $h=-1$ satisfies the equation. 

For $x_1,x_2$ inside region $\mathcal D'$ (but not on $\mathcal{L}$), $\ctwo(x_1,x_2)=0$ and $\ctwo(x_1,0)=-1/\nvac$ and the OZ equation becomes
\begin{equation}
    h(x_1,x_2) = - \frac{1-\nvac}{\nvac} h(0,x_2) = \frac{1-\nvac}{\nvac} \;  ,
\end{equation}
since $x_1=0,x_2 < \sigma$ are inside $\mathcal D$ and $h(0,x_2)=-1$. Thus, the total correlation function has the same $1/\nvac$ singularity as $\ctwo$ but on the complementary domain $\mathcal D'$.

Finally, outside the support $\mathcal D \cup \mathcal D'$ one obtains $h=0$. 

\section{The second-order direct correlation function for a single density peak}
\label{app:dcf_peak}

\begin{figure}
  \centering
  \includegraphics[width=\linewidth]{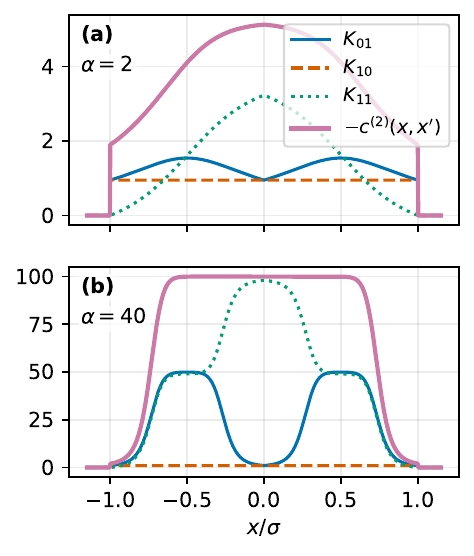}
  \caption{Decomposition of $-\ctwo(x,0)$ into the kernels $K_{01}$, $K_{10}$, and $K_{11}$. The mixed kernels sample the inverse free volume at exclusion-interval boundaries, while $K_{11}$ measures whether the overlap interval crosses a low-free-volume boundary layer.}
  \label{fig:c2-components}
\end{figure}

We consider a single Gaussian peak $\rho_{q\alpha}(x)$ and analyze the different kernel contributions $K_{ij}$ to $-\ctwo(x,x')$, see \Cref{eq:c21d,eq:Kij}. 
\Cref{fig:c2-components} shows these contributions along the cut $x'=0$ for a wide peak [$\alpha \sigma^2=2$, panel (a)] and a narrow peak [$\alpha \sigma^2=40$, panel (b)]. 
The contribution $K_{01}$ contains the inverse free volume $1/[2 A(x \pm R)]$ (shifted to the right and left). For the sharp peak it is visible that 
the integral of $K_{11}$ contains the same shifted terms $1/(2 A(x \pm R)$ plus a compensating term near $x=0$ with the magnitude $1/A(x)$. These terms add up to produce the plateau-like shape of $-\ctwo$ discussed in the main text.

\bibliography{literature}

\end{document}